\documentclass[sigconf]{acmart}

\AtBeginDocument{%
  \providecommand\BibTeX{{%
    \normalfont B\kern-0.5em{\scshape i\kern-0.25em b}\kern-0.8em\TeX}}}

\copyrightyear{2024}
\acmYear{2024}
\setcopyright{acmlicensed}

\acmConference[KDD '24]{Proceedings of the 30th ACM SIGKDD Conference on Knowledge Discovery and Data Mining }{August 25--29, 2024}{Barcelona, Spain.}

\acmBooktitle{Proceedings of the 30th ACM SIGKDD Conference on Knowledge Discovery and Data Mining (KDD '24), August 25--29, 2024, Barcelona, Spain}

\acmDOI{10.1145/3637528.3671953}

\acmISBN{979-8-4007-0490-1/24/08}

\usepackage{textcomp}
\usepackage{verbatim}
\usepackage{amsfonts}
\usepackage{amsmath}
\usepackage{array}
\usepackage{subfigure}
\usepackage{caption}
\usepackage{balance}
\usepackage{multicol}
\usepackage{multirow}
\usepackage{color}
\usepackage{epsfig}
\usepackage{natbib}
\usepackage{xspace}
\usepackage{booktabs}
\usepackage{bm}
\usepackage{enumitem}
\usepackage{diagbox}
\usepackage{url}
\usepackage{geometry}
\usepackage{changepage}
\usepackage{textcomp}
\usepackage{graphicx}
\usepackage{color}
\usepackage{bm}
  
\usepackage{amssymb}
\usepackage{algorithm}
\usepackage{algorithmic}
\usepackage{threeparttable}
\usepackage{mathrsfs}

\newcommand{\model}{\textit{Brant-X}\xspace}
\newcommand{\brantTwo}{Brant-2\xspace}
\newcommand{\brantOne}{Brant\xspace}

\newcommand{\vpara}[1]{\vspace{0.05in}\noindent\textbf{#1 }}
\newcommand{\vvpara}[1]{\vspace{0.06in}\noindent\textit{#1 }}

\begin{document}

\title{
$\vcenter{\hbox{\includegraphics[width=3ex,height=3ex]{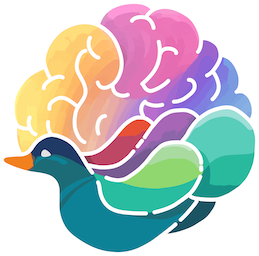}}}$
Brant-X: A Unified Physiological Signal Alignment Framework}

\author{Daoze Zhang}
\affiliation{%
  \institution{Zhejiang University}
  \city{Hangzhou}
  \country{China}
}
\email{zhangdz@zju.edu.cn}

\author{Zhizhang Yuan}
\affiliation{%
  \institution{Zhejiang University}
  \city{Hangzhou}
  \country{China}
}
\email{zhizhangyuan@zju.edu.cn}

\author{Junru Chen}
\affiliation{%
  \institution{Zhejiang University}
  \city{Hangzhou}
  \country{China}
}
\email{jrchen\_cali@zju.edu.cn}

\author{Kerui Chen}
\affiliation{%
  \institution{Zhejiang University}
  \city{Hangzhou}
  \country{China}
}
\email{chenkr@zju.edu.cn}

\author{Yang Yang}
\authornotemark[0]
\affiliation{%
  \institution{Zhejiang University}
  \city{Hangzhou}
  \country{China}
}
\email{yangya@zju.edu.cn}
\authornote{Corresponding author.}


\renewcommand{\shortauthors}{Daoze Zhang, Zhizhang Yuan, Junru Chen, Kerui Chen, and Yang Yang}





\begin{abstract}

Physiological signals serve as indispensable clues for understanding various physiological states of human bodies.
Most existing works have focused on a single type of physiological signals for a range of application scenarios. 
However, as the body is a holistic biological system, the inherent interconnection among various physiological data should not be neglected. 
In particular, given the brain's role as the control center for vital activities, \textit{electroencephalogram} (EEG) exhibits significant correlations with other physiological signals.  
Therefore, the correlation between EEG and other physiological signals holds potential to improve performance in various scenarios.
Nevertheless, achieving this goal is still constrained by several challenges: the scarcity of simultaneously collected physiological data, the differences in correlations between various signals, and the correlation differences between various tasks. 
To address these issues, we propose a unified physiological signal alignment framework, \model, to model the correlation between EEG and other signals. 
Our approach (1) employs the EEG foundation model 
to data-efficiently transfer the rich knowledge in EEG to other physiological signals, 
and (2) introduces the \textit{two-level alignment} to fully align the semantics of EEG and other signals from different semantic scales.   
In the experiments, \model achieves state-of-the-art performance compared with task-agnostic and task-specific baselines on various downstream tasks in diverse scenarios, including sleep stage classification, emotion recognition, \textit{freezing of gaits} detection, and eye movement communication. 
Moreover, the analysis on the arrhythmia detection task and the visualization in case study further illustrate the effectiveness of \model in the knowledge transfer from EEG to other physiological signals. 
The model's homepage is at \url{https://github.com/zjunet/Brant-X/}.

\end{abstract}

\maketitle

\section{Introduction} \label{sec:intro}

Physiological signals, as indispensable biomarkers, characterize the underlying complexities of the human body and encapsulate a wide range of critical information about an individual's health, with great significance for health monitoring, disease diagnosis, and treatment~\citep{muhammad2021comprehensive,he2022flexible}.
Among these, several key signals including 
\textit{electroencephalogram} (EEG), \textit{electrooculography} (EOG), \textit{electrocardiogram} (ECG) and \textit{electromyogram} (EMG), are especially essential in capturing primary physiological manifestations \citep{rim2020deep}. 
For instance, 
EEG signals, which record neural activity in the brain, have been utilized to study different stages of sleep and human emotions, aiding in diagnosing sleep-related disorders and emotional health issues~\citep{phan2022automatic}. 
Also, EOG signals, owing to their ability to monitor potential changes during eyeball movements, have proved instrumental in enabling communication for individuals living with \textit{neurodegenerative disorders}~\citep{chaudhary2016brain,tonin2020auditory}. 
Moreover, ECG signals, which record the fluctuation of the heart’s bio-electric activities, have been widely employed in investigations relating to cardiac health and diseases~\citep{xiao2023deep}.
Finally, EMG signals capture the electrical activity of human muscles, helping the diagnosis and rehabilitation training of neuromuscular diseases~\citep{al2023electromyography}. 
The applications of these physiological signals allow clinicians to monitor individual health in real-time and make data-driven decisions, holding far-reaching implications for many research fields like healthcare.  

\begin{figure}[h]
  \vspace{-1mm} 
  \centering
  \includegraphics[width=\linewidth / 100 * 100]{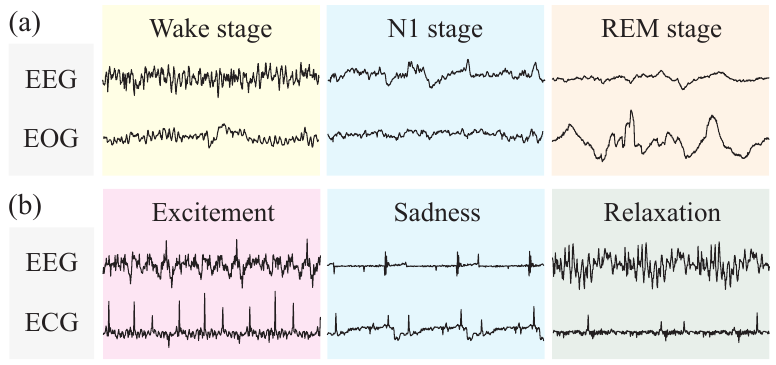}
  \vspace{-3mm} 
  \caption{Illustration of inherent correlations between EEG and other physiological signals. 
  \small 
  (a) 
  The waveform patterns in EEG and EOG vary with different sleep stages, especially with the REM stage, which is marked by rapid oscillations in EOG.
  (b) Excitement boosts heart beats in ECG with enhanced $\beta$ waves evident in EEG. Sadness slows heart rate and increases the brain activity in low-frequency $\alpha$ band. During relaxation, ECG presents a stable heart rate with heightened high-frequency EEG $\theta$ waves. 
  }
  \vspace{-3.5mm}
  \label{fig:sig_corr}
\end{figure}


Despite each physiological signal records the physiological conditions of its corresponding body part, it is worth noting that the body functions as an integrated biological system rather than some independent components~\citep{yi2019integration}. Thus, there exists an inherent interconnection among different physiological signals. Among these, given the brain's role as the epicenter for controlling vital activities, EEG exhibits \textit{significant correlations with synchronous physiological signals} from other body parts~\citep{kanwal2021internal}. 
Specifically, in some scenarios, since the information of single-type signal may be insufficient or noisy, ignoring this correlation can lead to great performance losses.  
Taking sleep staging as an example, as shown in Fig.~\ref{fig:sig_corr}(a), although EEG records different brainwaves in different stages, 
the rapid oscillations of EOG are particularly essential criteria for the \textit{rapid eye movement} (REM) stage.
Moreover, \citet{sharma2022automated} has also shown that introducing EOG signals can bring a relative improvement of 14.98\% in accuracy. 
Besides, the correlations between EEG and other signals also exists in other scenarios: 
\textbf{(1) EEG\&EOG}: 
For individuals with neurodegenerative disorders who can only express their thoughts and achieve interaction through eye movements, EEG and EOG can contribute to the development of assistive communication systems~\citep{tonin2020auditory}. 
\textbf{(2) EEG\&ECG}: During different emotional states in Fig.~\ref{fig:sig_corr}(b), brain signals and heartbeats consistently present different patterns, such that EEG and ECG can be utilized for emotion recognition~\citep{hasnul2021electrocardiogram}. 
\textbf{(3) EEG\&EMG}: Since the abrupt muscle rigidity (named \textit{freezing of gaits}, FoG) of Parkinson’s disease is related to a complex interplay between motor, cognitive and affective factors, EEG and EMG can be employed in FoG detection to enhance patient safety and quality of life~\citep{zhang2022multimodal}.
Hence, the correlations between EEG and other physiological signals (refered to as ``EXG'' in this paper, including EOG, ECG, and EMG) hold potential to improve performance in a variety of scenarios. 
Therefore, our work focuses on establishing an EEG-centric unified framework for modeling the correlation between EEG and EXG, which exploits the combined information of EEG and EXG to contribute to various application scenarios.  
However, current researches leave much to be explored in this direction, primarily due to the following challenges. 

\begin{figure*}
  \centering
  \includegraphics[width=\linewidth]{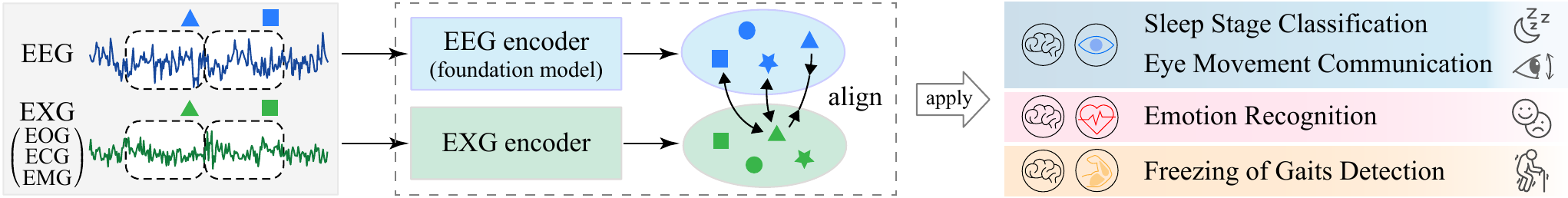}
  \vspace{-5mm}
  \caption{Overview of the physiological signal alignment framework \model.  
  \small 
  Firstly, based on the EEG foundation model, the EXG encoder is trained by the alignment between simultaneously collected EEG and EXG data. Then, the EEG and EXG encoders, capable of learning strong representations from EEG and EXG signals, are applied to various downstream tasks in diverse scenarios. 
  }
  \vspace{-3mm}
  \label{fig:framework}
\end{figure*}

From the viewpoint of data, \textbf{simultaneously collected EEG and EXG signals face a conspicuous lack of data.}
Due to the acquisition costs, ethical restrictions, and a lack of emphasis on the signal correlation in current machine learning research, the majority of physiological data records only a single type of signal, such as EEG datasets of several terabytes in size~\citep{harati2014tuh}. 
In contrast, available multi-type physiological datasets, which contain various physiological data collected simultaneously, are much smaller in scale, most being less than a few gigabytes. 
Therefore, the scarcity of simultaneously collected EEG and EXG data poses challenges in training a unified framework for modeling the correlation between EEG and EXG.

From a method perspective, \textbf{there exist significant inherent differences in correlations between EEG and different EXG signals. } 
Different types of physiological signals
differ greatly in their inherent properties such as amplitude and bandwidth~\citep{thakor2017biopotentials}. 
To satisfy the sampling theorem~\citep{shannon1949communication}, the huge gap in bandwidth further leads to differences in sampling rates.  
Specifically, due to the gap in bandwidth, the sampling rates for EOG, ECG, and EMG may vary respectively within the ranges of 50-100Hz, 250-500Hz, and 1000-2000Hz. 
These discrepancies are also evident in other features like typical waveforms and rhythmicity~\citep{thakor2017biopotentials}. 
The above factors result in vast inherent differences in correlations between EEG and different EXGs, posing a challenge to the unified modeling method of EEG-EXG correlation. 


From the viewpoint of task, \textbf{in different scenarios, various downstream tasks depend on different correlations even between EEG and the same EXG. }
Given that different application scenarios involve different physiological activities of body organs, different downstream tasks need to capture different correlations between EEG and even the same EXG. 
Specifically, 
since the physiological changes during sleep is relatively slow, in sleep staging task, the EEG-EOG correlation is required to capture on a scale up to 30sec, which is defined as a sleep stage~\citep{berry2017aasm}. 
In contrast, in eye movement communication task, eyeball movements may occur in less than 1sec, depending on different EEG-EOG correlation from sleep staging~\citep{jaramillo2021dataset}. 
Therefore, it is challenging to capture different EEG-EXG correlations for various downstream scenarios.  


To tackle the above issues, we propose a contrastive-learning-based framework named \model, to efficiently align EEG and EXG signals from different semantic scales for the modeling of correlation between EEG and EXG. 
To address the scarcity of simultaneously collected EEG and EXG data, 
our intuitive idea is to use models trained with a large amount of EEG data to empower the representation learning on EXG signals. 
Inspired by large language models that are widely applied in other research fields like computer vision~\citep{wang2023visionllm,menon2022visual},  
we employ the EEG foundation model \brantTwo~\citep{zhang2023brant,yuan2024brant}, which is pre-trained on 4TB brain signal data and contains 1B parameters.  
Based on this, we summarize existing public multi-type physiological datasets\footnote{
For the details about the review of public multi-type physiological datasets, please refer to \url{https://github.com/zjunet/Brant-X/}
}, to perform data-efficient knowledge transfer from EEG to EXG. 
Observing that the gaps between tasks primarily stem from the differences in \textit{semantic scales} of correlation, to address the gaps among various signals and tasks,
we introduce the \textit{two-level alignment} that aligns the semantics of EEG and EXG at both patch- and sequence-level.
The patch-level alignment overcomes finer inherent differences and captures EEG-EXG correlation at a smaller semantic scale, while the sequence-level one aligns coarser differences and captures the correlation at a larger scale. 
Moreover, we adopt the \textit{sampling augmentation} to enhance model robustness to different sampling rates.  
Using the above methods, data and model resources in EEG are extended to empower the research on other physiological signals, paving a new avenue to model the correlations between various physiological signals.


To validate the effectiveness of \model, extensive experiments show that \model achieves SOTA performance on various downstream tasks across diverse scenarios involving EEG and EXG signals, 
including sleep stage classification, emotion recognition, freezing of gaits detection, and eye movement communication. 
The analysis on the arrhythmia detection task and the visualization in case study further demonstrate that \model can effectively transfer the knowledge from EEG to EXG signals through alignment. 
Overall, our key contributions comprise: 
\begin{itemize}[leftmargin=*]
    \item We are the first to design a unified EEG-centric alignment framework to model the correlations between EEG and other physiological signals, which can be applied to various scenarios. 
    \item Based on the EEG foundation model, we adopt the two-level alignment for data-efficient knowledge transfer from EEG to EXG signals, which combines the semantics of EEG and EXG to jointly improve the performance on downstream tasks.
    \item We validate \model through extensive experiments on multiple downstream tasks involving various physiological signals.
    Moreover, the analysis and visualization illustrate the effectiveness of \model in knowledge transfer from EEG to EXG. 
\end{itemize}

\section{Proposed Method}

\begin{figure*}[ht]
  \centering
  \includegraphics[width=\linewidth / 100 * 92]{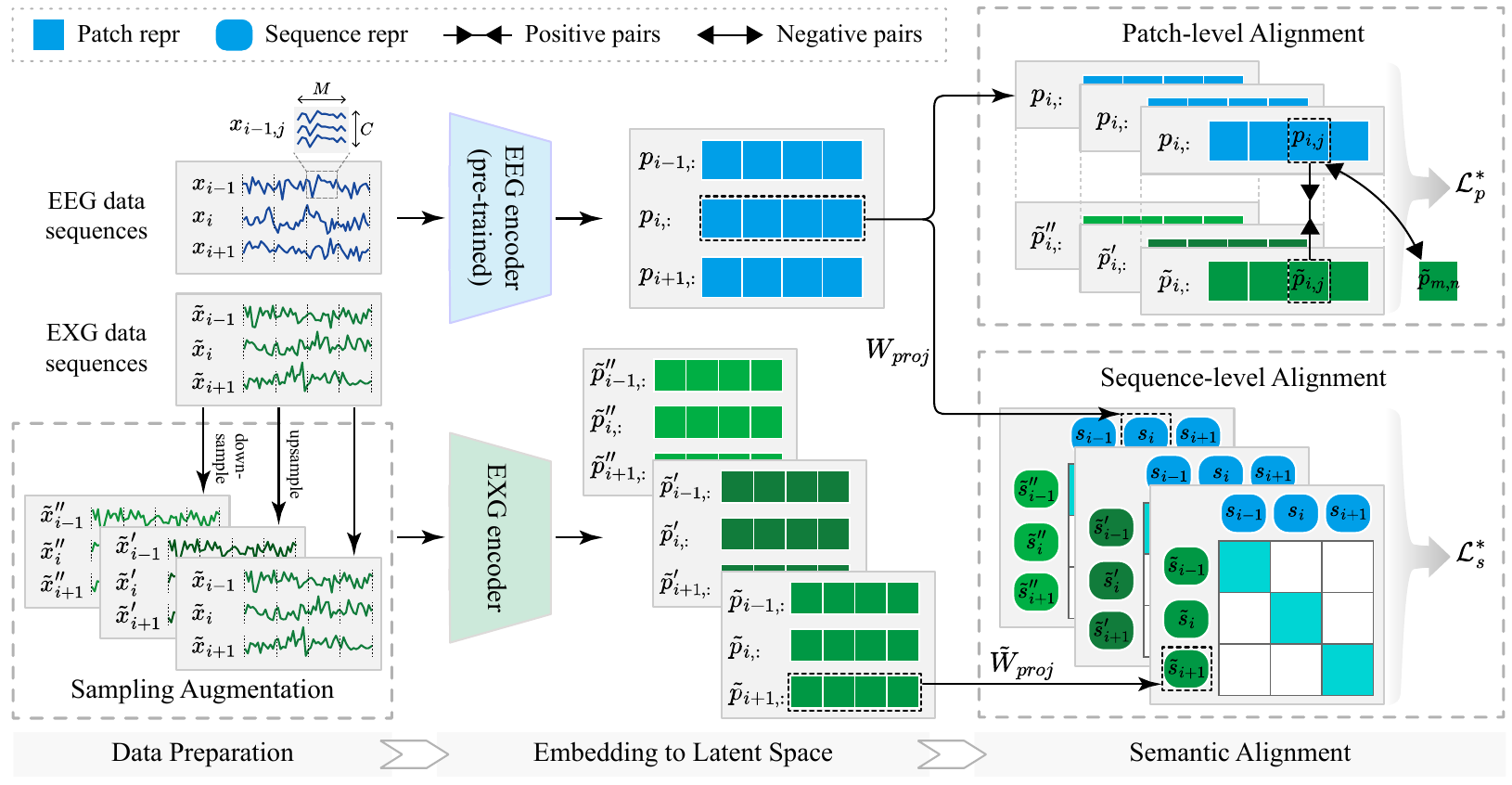}
  \caption{Architecture of \model.  
  \small 
  In the data preparation stage, EXG data are upsampled and downsampled for data augmentation. Then, EEG and EXG data are fed into the EEG encoder and EXG encoder respectively to obtain the representations of  data patches. Finally, we align the simultaneously collected EEG and EXG patches and the corresponding EEG and EXG sequences by two-level alignment. 
  }
  \label{fig:align}
\end{figure*}

In this section, we introduce the technical details of the proposed framework \model. 
Specifically, as shown in the upper left part of Fig.~\ref{fig:align}, we first split the EEG and EXG sequences into continuous data patches. 
Then, considering the variance in sampling rates between physiological signals in different scenarios, we adopt the sampling augmentation (lower left corner of Fig.~\ref{fig:align}) to enhance the model's robustness to changes in sampling rates. 
As shown in middle part of Fig.~\ref{fig:align}, the EEG patches and EXG patches, along with the augmented patches, are fed into the EEG and EXG encoder, respectively, to acquire the representation of each data patch. 
Here we employ the EEG foundation model \brantTwo as the EEG encoder of our framework (details in Sec.~\ref{sec:brant2}).
During the unsupervised training process, we propose the two-level alignment (right part of Fig.~\ref{fig:align}), which aligns the simultaneously collected patches and sequences at both patch- and sequence-level. 
After the unsupervised alignment, the representations of EEG and EXG data output by the two encoders will be aggregated via the attention mechanism for various tasks in diverse scenarios. 

\subsection{Problem Formulation}

First, we formalize the definitions of the four downstream tasks where the experiments are conducted. 
The collection of physiological signals relies on signal collection pads, referred to as \textit{electrodes}, distributed on the body part to be monitored. Multiple electrodes simultaneously record the bioelectric activity of the corresponding organs, generating a multi-\textit{channels} time series. 
Formally, given $S$ EEG signal sequences $\{ \boldsymbol{x}_i \}_{i=0}^{S-1}$ that correspond to $S$ physiological processes, each data sequence $\boldsymbol{x}_i \in \mathbb{R}^{C \times L}$ includes $C$ channels with a length of $L$ timestamps. The simultaneously collected EXG signals, including $\tilde{C}$ channels, are denoted as $\{ \boldsymbol{\tilde{x}}_i \}_{i=0}^{S-1}$, where $\boldsymbol{\tilde{x}}_i \in \mathbb{R}^{\tilde{C} \times L}$. 
According to different tasks or scenarios, each multi-type sequence $\{ \boldsymbol{x}_i, \boldsymbol{\tilde{x}}_i\}$ is annotated with a label $y_i \in \mathbb{R}$ by professional physicians. 
Based on the above, our research problems can be defined as: 
\begin{definition}
  Given EEG data sequences $\{ \boldsymbol{x}_i \}_{i=0}^{S-1}$ and EXG data sequences $\{ \boldsymbol{\tilde{x}}_i \}_{i=0}^{S-1}$, with the corresponding labels $\{ y_i \}_{i=0}^{S-1}$, 
  the aim is to classify each multi-type sequence $\{ \boldsymbol{x}_i, \boldsymbol{\tilde{x}}_i\}$ to determine which class it belongs to.
\end{definition}

\subsection{Foundation Models for EEG} \label{sec:brant2}


Due to the scarcity of simultaneously collected physiological data, it is challenging to build a unified framework for the modeling of correlations between EEG and EXG. 
To address this issue,  
we adopt the brain signal foundation model to perform data-efficient knowledge transfer from EEG to EXG. 
To the best of our knowledge, only the series of works named \brantOne currently serves as open-source foundation models on brain signals, including Brant and Brant-2. 
Specifically, ~\citet{zhang2023brant} provide the first off-the-shelf foundation model named \brantOne for intracranial EEG (iEEG)\footnote{
Compared to EEG signals recorded on the surface of the scalp, iEEG relies on implanted electrodes to measure deep brain activity. However, due to the required cranial surgery and ethical restrictions, the application of iEEG is not as widespread as EEG. 
} signals, which contains 500M parameters pre-trained on 1.01TB iEEG data. 
Based on \brantOne, \citet{yuan2024brant} propose the foundation model for brain signals named \brantTwo. It consists of over 1B parameters and is pre-trained on as much as nearly 4TB mixed data (with 2.3TB iEEG data from 26 subjects and 1.6TB EEG data from about 15,000 subjects). 
Our choice to use \brantTwo as the EEG encoder in our framework was two-fold. Firstly, it is pre-trained on a large corpus of brain signal data and can learn powerful representations from EEG signals. 
More importantly, it 
uses pre-training data with different sampling rates, resulting in a heightened level of robustness towards changes in sampling rates.

\subsection{Overall Architecture}

\vpara{Patching.}
Given that physiological data are bioelectric signals,
the semantic information of the physiological states can only be collectively expressed with multiple sampling points, rather than a single one. 
Therefore, we split a whole data sequence into several consecutive patches to aggregate semantic information within patches and reduce computation demand~\cite{nie2022time}. 

Formally, as shown in the upper left part of Fig.~\ref{fig:align}, given the $i$-th multi-channel EEG data sequence $\boldsymbol{x}_i \in \mathbb{R}^{C \times L}$ where $C$ denotes the number of EEG channels and $L$ denotes the number of timestamps (length of the sequence), we split $\boldsymbol{x}_i$ with length $M$ to generate a set of non-overlapping patches $\{ \boldsymbol{x}_{i,j} \}_{j=0}^{P-1}$, where $\boldsymbol{x}_{i,j} \in \mathbb{R}^{C \times M}$ and $P=\lfloor L/M
 \rfloor$ is the number of patches in this sequence. 
For the EXG data, we apply the same patching process as above, and the symbols are also similar. 
Specifically, we use $\boldsymbol{\tilde{x}}_i \in \mathbb{R}^{\tilde{C} \times \tilde{L}}$ to denote the $i$-th EXG sequence, where $\tilde{C}$ is the number of EXG channels and $\tilde{L}$ is the sequence length. 
Also, $\{ \boldsymbol{\tilde{x}}_{i,j} \}_{j=0}^{P-1}$ denotes the set of patches,
where $\boldsymbol{\tilde{x}}_{i,j} \in \mathbb{R}^{\tilde{C} \times M}$. 

\vpara{Sampling Augmentation. }
Considering that different physiological signals exhibit large differences in sampling rates, models that learn representations from physiological data must be sufficiently robust to changes in sampling rates. 
For EEG, as presented in Sec.~\ref{sec:brant2}, \brantTwo utilize pre-training data at various sampling rates, making it fairly robust to changes in sampling rates. Hence, serving as the EEG encoder of our framework, it is capable of handling differences in the sampling rate of EEG data. 

For the EXG signals, to address the issue of various sampling rates, 
we adopt sampling augmentation to enhance the model’s robustness to changes in sampling rate. 
Specifically, as shown in the lower left corner of Fig.~\ref{fig:align}, we both upsample the original data to twice its original rate and downsample it to half, producing two sets of augmented data with different sampling rates. 
Formally, given the original EXG data patches $\{ \boldsymbol{\tilde{x}}_{i,j} \}_{j=0}^{P-1} $, we upsample the data to twice its sampling rate, generating the upsampled data patches $\{ \boldsymbol{\tilde{x}}_{i,j}' \}_{j=0}^{P-1}$ where $\boldsymbol{\tilde{x}}_{i,j}' \in \mathbb{R}^{\tilde{C} \times 2M}$. Similarly, the original patches are also downsampled to half the sampling rate, thus obtaining the downsampled data patches $\{ \boldsymbol{\tilde{x}}_{i,j}'' \}_{j=0}^{P-1}$ where $ \boldsymbol{\tilde{x}}_{i,j}'' \in \mathbb{R}^{\tilde{C} \times \lfloor M/2 \rfloor}$. 

In subsequent representation learning and semantic alignment sections, the original data $\boldsymbol{\tilde{x}}$, along with the upsampled data $\boldsymbol{\tilde{x}}'$ and downsampled data $\boldsymbol{\tilde{x}}''$, will be fed into the EXG encoder for model learning purposes.

\vpara{Embedding to Latent Space. }
For EEG data, as shown in the middle part of Fig.~\ref{fig:align}, we feed it directly into the pre-trained EEG encoder (details in Sec.~\ref{sec:brant2}) to obtain the EEG representation. 
Formally, $P$ consecutive patches $\{ \boldsymbol{x}_{i,j} \}_{j=0}^{P-1}$ from the $i$-th EEG data sequence $\boldsymbol{x}_i$ will be input into the EEG encoder, yielding the representations $\{ \boldsymbol{p}_{i,j} \}_{j=0}^{P-1}$ of these patches, 
where $\boldsymbol{p}_{i,j} \in \mathbb{R}^{D_p}$ denotes the representation of the $j$-th patch from the $i$-th sequence of EEG data, and $D_p$ denotes the dimension of patch representations. 

When it comes to EXG data, it will be fed into the EXG encoder to obtain its representation.  
Formally, all the patches $\{ \boldsymbol{\tilde{x}}_{i,j} \}_{j=0}^{P-1}$ from the $i$-th EXG data sequence $\boldsymbol{\tilde{x}}_i$ are input into the EXG encoder, generating their representations $\{ \boldsymbol{\tilde{p}}_{i,j} \}_{j=0}^{P-1}$, where $\boldsymbol{\tilde{p}}_{i,j} \in \mathbb{R}^{D_p}$. 
Given that the focus of our work is the alignment framework, the specific architecture of the EXG encoder can be flexible. For the technical details of the EXG encoder used in this paper, please refer to App.~\ref{app:exg_encoder}.

Similarly, the upsampled EXG patches $\{ \boldsymbol{\tilde{x}}_{i,j}' \}_{j=0}^{P-1}$ and the downsampled patches $\{ \boldsymbol{\tilde{x}}_{i,j}'' \}_{j=0}^{P-1}$ undergo the same process, obtaining the representations $\{ \boldsymbol{\tilde{p}}_{i,j}' \}_{j=0}^{P-1}$ and $\{ \boldsymbol{\tilde{p}}_{i,j}'' \}_{j=0}^{P-1}$ of augmented EXG data.

\subsection{Two-level Alignment}


We adopt two-level alignment that fully aligns the semantics of EEG and EXG signals at patch- and sequence-level, to overcome inherent differences and capture the correlation between EEG and EXG at different semantic scales. 

\vpara{Patch-level Alignment. }
At a finer grain, we align EEG and EXG data at patch-level by placing the simultaneous EEG and EXG patches close together in the latent space, while mapping unrelated patches further apart. 
As shown in the upper right part of Fig.~\ref{fig:align}, since our EEG encoder is pre-trained on a large amount of data (Sec.~\ref{sec:brant2}), it is reasonable to believe it can output representative representations of EEG patches. 
Therefore, we set the EEG representation $\boldsymbol{p}_{i,j}$ as the anchor. The anchor $\boldsymbol{p}_{i,j}$ and the simultaneously collected EXG patch $\boldsymbol{\tilde{p}}_{i,j}$ are set as the positive sample pair. 
Negative samples are randomly selected from the representations $\{ \boldsymbol{\tilde{p}}_{m} \}_{m \neq i}$ from other EXG data sequences. 
It is noteworthy that, contrary to the sequence-level alignment described later, we can't randomly select the representations $\{ \boldsymbol{\tilde{p}}_{i,n} \}_{n \neq j}$ from the EXG sequence $\boldsymbol{\tilde{p}}_{i}$ as negative samples. This is because these representations $\{ \boldsymbol{\tilde{p}}_{i,n} \}_{n \neq j}$ and the anchor originate from the same physiological process and may have a temporal dependency between them. 
Formally, for the anchor $\boldsymbol{p}_{i,j}$, the negative sample set $Z^p_{i,j}$ is randomly sampled from all the negative samples $\{ \boldsymbol{\tilde{p}}^{}_{m,n} | m \neq i, n=0,...,P-1 \}$. 
The InfoNCE~\citep{oord2018representation} loss is applied to retain the maximum mutual information between positive pairs: 
\begin{equation}
    \mathcal{L}_p = 
    \frac{1}{S P} 
    \sum_{i} \sum_{j} 
    -\text{log} 
    \frac{\text{exp}(
    \boldsymbol{p}_{i,j}^\intercal \boldsymbol{\tilde{p}}^{}_{i,j} / t_p
    )}
    {
    \sum_{\boldsymbol{\tilde{p}}^{}_{m,n} \in Z^p_{i,j} } 
    \text{exp}(
    \boldsymbol{p}_{i,j}^\intercal \boldsymbol{\tilde{p}}^{}_{m,n} / t_p
    )
    } ,
\end{equation}
where $t_p$ denotes the temperature hyperparameter to adjust scale, and $\mathcal{L}_p$ denotes the InfoNCE loss between EEG and original EXG data in patch-level alignment.

Similarly, the same alignment process would also exist between the EEG data and the two sets of augmented EXG data. 
These two losses are denoted as $\mathcal{L}_p'$ and $\mathcal{L}_p''$, respectively. Overall, the optimization objective of patch-level alignment is given by: 
\begin{equation}
    \mathcal{L}_p^\ast = 
    \mathcal{L}_p + \mathcal{L}_p' + \mathcal{L}_p'' .
\end{equation}

\vpara{Sequence-level Alignment. }
At a coarser granularity level, we employ sequence-level alignment to align the corresponding sequence in the latent space. 
To aggregate the representations of patches from a data sequence, we firstly perform a linear projection $W_{proj} \in \mathbb{R}^{D_s \times P D_p}$ on all patch representations $\boldsymbol{p}_{i,:} \in \mathbb{R}^{P \times D_p}$ from sequence $\boldsymbol{x}_{i}$, thus obtaining the sequence representation $\boldsymbol{s}_{i} \in \mathbb{R}^{D_s}$, 
where $D_s$ denotes the dimension of sequence representations:
\begin{equation}
    \boldsymbol{s}_{i} = W_{proj} \left(
        \text{Flatten}(\boldsymbol{p}_{i,:})
    \right).
\end{equation}
This linear projection is applied similarly for EXG data $\boldsymbol{\tilde{p}}_{i,:}$ and the augmented data $\boldsymbol{\tilde{p}}_{i,:}', \boldsymbol{\tilde{p}}_{i,:}''$ as well, yielding the sequence representations $\boldsymbol{\tilde{s}}_{i}$, $\boldsymbol{\tilde{s}}_{i}'$ and $\boldsymbol{\tilde{s}}_{i}''$ respectively. 

After obtaining the sequence representations, we set the representations of simultaneously collected EEG and EXG sequences ($\boldsymbol{s}_{i}$ and $\boldsymbol{\tilde{s}}_{i}$) as positive sample pairs, while all other sequence pairs are set as negative pairs. 
Formally, the negative sample set $Z^s_i$ of sequence $\boldsymbol{s}_{i}$ is randomly sampled from all the negative samples $\{ \boldsymbol{\tilde{s}}^{}_{m} | m \neq i \}$. 
The sequence-level InfoNCE loss $\mathcal{L}_s$ for the EEG and the original EXG data can be given as follows: 
\begin{equation}
    \mathcal{L}_s = 
    \frac{1}{S} 
    \sum_{i}
    -\text{log} 
    \frac{\text{exp}(
    \boldsymbol{s}_{i}^\intercal \boldsymbol{\tilde{s}}^{}_{i} / t_s
    )}
    {
    \sum_{ \boldsymbol{\tilde{s}}^{}_{m} \in Z^s_i }
    \text{exp}(
    \boldsymbol{s}_{i}^\intercal \boldsymbol{\tilde{s}}^{}_{m} / t_s
    )
    } ,
\end{equation}
where $t_s$ denotes the temperature hyperparameter. As shown in the bottom right part of Fig.~\ref{fig:align}, following the common practice in CLIP~\citep{radford2021learning}, we adopt a similarity matrix to optimize this objective. 

Likewise, we carry out the same alignment process between EEG and augmented EXG data, resulting in two losses $\mathcal{L}_s'$ and $\mathcal{L}_s''$ in the same form. The overall loss in sequence-level alignment is: 
\begin{equation}
    \mathcal{L}_s^\ast = 
    \mathcal{L}_s + \mathcal{L}_s' + \mathcal{L}_s'' .
\end{equation}

Finally, the objective of joint optimization is obtained by adding the patch-level and sequence-level alignment losses $\mathcal{L}_p^\ast$ and $\mathcal{L}_s^\ast$.

\section{Experiment}

\subsection{Experimental Setup} 

\vpara{Alignment. } 
To align the simultaneously recorded EEG and arbitrary EXG data, the training data used for unsupervised alignment is collectively assembled from three datasets: CAP~\citep{terzano2001atlas}, ISRUC~\citep{khalighi2016isruc}, and HMC~\citep{alvarez2021inter}, which include EEG, EOG, ECG, and EMG signals. 
Overall, the alignment training data includes 359 recordings from 267 subjects. 
The alignment is performed on a Linux system with 2 CPUs (AMD EPYC9654 96-Core Processor) and 2 GPUs (NVIDIA Tesla A100 80G). 
The learning rate of EEG encoder is set as $1 \times 10^{-5}$ for finetuning, while the EXG encoder is trained with a higher learning rate of $3 \times 10^{-4}$.

\vpara{Downstream Tasks. } 
Here we introduce the four downstream tasks used to validate the effectiveness of our \model, along with the datasets, setups and and evaluation metrics. 

\vvpara{$\bullet$ Sleep Stage Classification. }
In sleep health research, sleep staging refines human understanding of sleep states and patterns, which holds significance for the prevention and diagnosis of sleep-related diseases~\citep{phan2022automatic}. 
According to the American Academy of Sleep Medicine (AASM) manual~\citep{berry2017aasm}, sleep occurs in five stages: wake, N1, N2, N3, and REM. Among these, N1 to N3 are \textit{non-rapid eye movement} sleep, with each stage leading to progressively deeper sleep. 
Hence, sleep stage classification is a five-class classification problem. 

As for the dataset, the Sleep-EDF datasets~\citep{kemp2000analysis} are very popular in sleep staging researches. The Sleep-EDF-78 dataset contains 153 whole-night  polysomnographic sleep recordings from sleep cassette studies, containing 100Hz EEG and EOG data from 78 subjects aged 25-101 years (37 males and 41 females). Data are segmented into 30sec epochs and manually annotated by experts. The Sleep-EDF-20 dataset, which contains 39 recordings from 20 subjects, is also used in our study to facilitate the comparison with the existing methods. 

The experiment is conducted on EEG and EOG signals in a subject-independent setting. 
We divide the subjects into training, validation, and test sets in a 3:1:1 ratio. The experiments are repeated on all subjects to obtain overall results. The evaluation metrics include accuracy, sensitivity, specificity, macro F1 score, and Cohen's kappa $\kappa$.

\vvpara{$\bullet$ Emotion Recognition. }
Automatic emotion recognition has made a remarkable entry in the domain of biomedical, brain-computer interface, smart environment, safe driving and so on~\citep{kamble2023comprehensive}.  
Emotions are categorized into two types: 
(1) discrete emotions like joy, fear and sadness; and 
(2) multi-dimensional emotions on three emotion dimensions: arousal, valence, and dominance dimensions.  
Existing works~\citep{katsigiannis2017dreamer,liu2020multi,lin2023eeg,sun2022dual,tao2020eeg} mainly focus on the recognition of multi-dimensional emotions, so the task can be regarded as three independent binary classification problems: low/high valence, low/high arousal and low/high dominance. 

The DREAMER dataset~\citep{katsigiannis2017dreamer} is used to conduct experiments on emotion recognition task. It contains EEG (128Hz) and ECG (256Hz) data of 23 subjects (14 males and 9 females) when they are watching 18 film clips. Each film clip has an average length of 199s, which is thought to be sufficient for eliciting single emotion.  After watching a film clip, emotion statuses are labeled as low or high on the three emotion dimensions, serving as the labels for emotion recognition. 

The experiment in this task is conducted on EEG and ECG signals in a subject-independent setting. 
We split subjects into training, validation, and test sets in a 3:1:1 ratio and repeat the experiments on all subjects. 
The evaluation metrics are mainly accuracy~\cite{liu2020multi,lin2023eeg,tao2020eeg}, with some studies~\citep{katsigiannis2017dreamer,sun2022dual} also including the F1 score and the AUC of precision-recall curve.

\def\one #1 {\textbf{#1}}
\def\two #1 {\underline{#1}~}
\def\thr #1 {#1*}
\def \s   #1 {\footnotesize $\pm$#1}

\begin{table*}[ht]
    \caption{Average performance on the sleep stage classification task.}
    \renewcommand{\arraystretch}{0.95}  
    \vspace{-2mm}
    \setlength\tabcolsep{2.5pt}
    \centering
    \label{tab:exp_sleep}

    \begin{tabular}{lllllllllll}
    \toprule
    \multicolumn{1}{c}{\multirow{2}{*}{\diagbox{Methods}{Metrics}}}
     & \multicolumn{5}{c}{Sleep-EDF-20}                                & \multicolumn{5}{c}{Sleep-EDF-78}                                \\
    \cmidrule(lr){2-6} \cmidrule(lr){7-11} 
                      & \multicolumn{1}{c}{Acc.} & \multicolumn{1}{c}{Sens.} & \multicolumn{1}{c}{Spec.} & \multicolumn{1}{c}{Macro F1} & \multicolumn{1}{c}{Kappa} & \multicolumn{1}{c}{Acc.} & \multicolumn{1}{c}{Sens.} & \multicolumn{1}{c}{Spec.} & \multicolumn{1}{c}{Macro F1} & \multicolumn{1}{c}{Kappa}     \\
    \midrule
    TF-C~\citep{zhang2022self}                    & 55.42 \s1.39 & 31.52 \s1.09 & 86.07 \s0.39 & 26.04 \s0.21 & 30.74 \s1.52 & 53.90 \s4.03 & 31.35 \s2.40 & 85.80 \s1.34 & 26.00 \s2.09 & 29.32 \s6.43 \\
    SimMTM~\citep{dong2023simmtm}                 & 66.91 \s1.89 & 53.47 \s1.58 & 90.61 \s1.63 & 53.21 \s1.95 & 53.25 \s2.02 & 63.06 \s2.67 & 59.12 \s3.88 & 91.21 \s1.56 & 57.07 \s2.13 & 53.07 \s3.42 \\
    OneFitsAll~\citep{zhou2023onefitsall}         & 72.60 \s1.51 & 63.50 \s8.36 & 92.76 \s1.12 & 61.61 \s5.80 & 61.81 \s3.50 & 68.50 \s2.19 & 56.58 \s4.16 & 91.34 \s0.86 & 54.24 \s1.96 & 55.21 \s3.07 \\
    Time-LLM~\citep{jin2023time} & 80.31 \s2.63          & 76.53 \s3.15          & 94.53 \s2.95          & 71.64 \s3.02          & 70.22 \s2.84          & 78.08 \s2.96          & 67.44 \s3.73          & 94.13 \s3.01          & 66.09 \s3.25          & 68.04 \s3.14          \\
    MiniRocket~\citep{dempster2021minirocket} & 81.60 \s1.55  & 72.63 \s1.80 & 95.15 \s1.12 & 72.82 \s2.01 & 72.79 \s1.96 & 78.36 \s1.93 & 69.76 \s2.44 & 94.08 \s1.76 & 70.18 \s2.35 & 69.46 \s2.46 \\

    \midrule
    TinySleepNet~\citep{supratak2020tinysleepnet} & \two83.64 \s2.31 & \one81.60 \s2.60 & \two96.05 \s2.08 & \two77.54 \s2.55 & \two77.63 \s2.29 & \one83.49 \s2.24 & \thr80.25 \s2.65 & \one96.02 \s2.11 & \two76.64 \s2.61 & \two76.41 \s2.59 \\
    XSleepNet~\citep{phan2021xsleepnet}           & 80.93 \s2.34 & 75.78 \s2.21 & 94.79 \s2.54 & \thr76.71 \s2.59 & 74.31 \s2.32 & \thr81.83 \s2.30 & \two80.50 \s2.28 & \thr95.74 \s2.58 & \thr75.28 \s2.66 & \thr75.44 \s2.37 \\
    L-SeqSleepNet~\citep{phan2023seqsleepnet}     & \thr82.90 \s2.12 & 78.42 \s2.25 & \thr95.86 \s2.00 & 74.90 \s2.22 & 76.47 \s2.24 & 80.84 \s2.18 & 72.75 \s2.54 & 95.19 \s2.34 & 72.67 \s2.38 & 74.94 \s2.51 \\ 
    SleepHGNN~\citep{jia2023exploiting}           & 81.15 \s1.96 & 74.23 \s2.10 & 94.93 \s1.96 & 72.88 \s2.17 & 73.35 \s2.16 & 77.35 \s2.13 & 69.94 \s2.48 & 94.04 \s2.02 & 69.56 \s2.39 & 68.65 \s2.41 \\
    SleepKD~\citep{liang2023teacher}  & 82.44 \s2.40          & 78.20 \s2.54          & 94.78 \s2.34          & 74.11 \s2.72          & \thr76.87 \s2.63          & 80.19 \s2.85          & 72.95 \s2.88          & 94.95 \s2.69          & 72.65 \s2.84          & 74.86 \s2.93          \\
    SleepDG~\citep{wang2024generalizable}  & 81.92 \s2.27          & \thr79.12 \s2.35          & 95.75 \s2.68          & 74.74 \s2.53          & 76.43 \s2.47          & 79.95 \s2.42          & 73.31 \s2.41          & 93.57 \s2.63          & 72.21 \s2.59          & 74.16 \s2.68          \\

    \midrule
    \model           & \one84.58 \s1.98 & \two80.18 \s2.23 & \one96.36 \s1.89 & \one77.63 \s2.13 & \one79.29 \s2.18 & \two82.84 \s2.21 & \one81.85 \s2.42 & \two95.91 \s2.08 & \one77.04 \s2.30 & \one76.67 \s2.49 \\
    \bottomrule
    \end{tabular}
    
    \vspace{-2mm}
\end{table*}

\vvpara{$\bullet$ Freezing of Gaits Detection. }
FoG, which refers to the interruption of the motion caused by the brain’s incompetence to deal with concurrent cognitive and motor request, affects about 50\%-80\% of Parkinson’s disease patients as one of the severest manifestations. Thus, accurate detection of FoG can significantly improve patients’ life quality and promote personalized treatment~\citep{zhang2022multimodal}. The FoG detection task is a binary classification problem, that is, determining whether FoG appears during a walking process. 

The FoG dataset~\citep{zhang2022multimodal} is used in this work, which includes EEG and EMG signals (1000Hz) collected from 12 Parkinson's disease patients (6 males and 6 females) aged 57-81 years with disease durations between 1 and 20 years. 
The valid data lasts for 3h42min, including 2h14min of normal gait and 1h28min of freezing of gait, labeled by two qualified physicians.  

The experiment in this task is conducted on EEG and EMG. 
The training, validation, and test data are randomly split in a 3:1:1 ratio. 
We also repeat the experiments to obtain the overall results. 
As a classification problem, the evaluation metrics used for this task are accuracy, precision, recall and F1 score.

\vvpara{$\bullet$ Eye Movement Communication. }
Due to paralysis caused by neurodegenerative disorders like \textit{amyotrophic lateral sclerosis} (ALS), many patients lost almost all their communication abilities~\citep{jaramillo2021dataset}, and only have remnant oculomotor control to form words, phrases, and sentences using a speller system~\citep{tonin2020auditory}.
The speller system works on a binary principle where the patient responds to auditory questions by moving their eyes to say ``yes'' and not moving the eyes for ``no''. Therefore, the eye movement communication task is also a binary classification problem (yes or no). 

The dataset published by~\citet{jaramillo2021dataset} is used for the eye movement communication experiment. The dataset contains EEG and EOG data (500Hz) recorded from four patients suffering from ALS. 
Data are recorded during 2-10 visits, each visit consisting of an average of 3.22 days with 5.57 sessions recorded per day. Due to the inconsistency in EOG channels across different files in the dataset, we exclude files lacking specific EOG channels to conduct the experiment. 

The experiment in this task is conducted using EEG and EOG. 
Experiments are conducted on training, validation, and test data split 8:1:1 and are repeated on all data files. 
The evaluation metrics are accuracy, precision, recall and F1 score.  

As for data pre-processing, for the three tasks except eye movement classification, we did not perform filtering or other processing, directly using the preprocessed data of the original datasets.
For the eye movement dataset, as the publisher didn't filter, we applied 45Hz low-pass filtering and z-score normalization.

\vpara{Baselines. } 
As a unified unsupervised alignment framework for physiological signal modeling, we compare \model with the advanced self-supervised or unsupervised methods designed for general time series on all the downstream tasks, including TF-C~\citep{zhang2022self} and SimMTM~\citep{dong2023simmtm}. 
Also, to compare \model with the methods that performs time series classification based on pre-trained language models, we set OneFitsAll~\citep{zhou2023onefitsall} and Time-LLM~\citep{jin2023time} as a baseline. 
As for the supervised methods, MiniRocket~\citep{dempster2021minirocket} is selected as our baseline due to its efficiency and versatility.
Furthermore, we compare \model with the SOTA methods that are specially designed for each task, to demonstrate the effectiveness of \model in various scenarios. These task-specific or signal-specific supervised methods includes: 
(1) TinySleepNet~\citep{supratak2020tinysleepnet}, XSleepNet~\citep{phan2021xsleepnet}, L-SeqSleepNet~\citep{phan2023seqsleepnet}, SleepHGNN~\citep{jia2023exploiting},
SleepKD~\citep{liang2023teacher}, and
SleepDG~\citep{wang2024generalizable} for sleep stage classification; 
(2) MLF-CapsNet~\citep{liu2020multi}, EEG-Conformer~\citep{song2022eeg}, \citet{lin2023eeg} and \citet{wang2023novel} for emotion recognition; 
(3) ~\citet{aly2023bio}, ~\citet{batool2022movement} and ~\citet{goel2023ensemble} for freezing of gaits detection; and 
(4) eyeSay~\citep{zou2021eyesay}, ~\citet{adama2021yes} and ~\citet{hossieny2022developing} for eye movement communication. More details about these baselines are given in App.~\ref{app:baseline}.

\begin{figure}[h]
  \centering
  \includegraphics[width=\linewidth / 10 * 10] {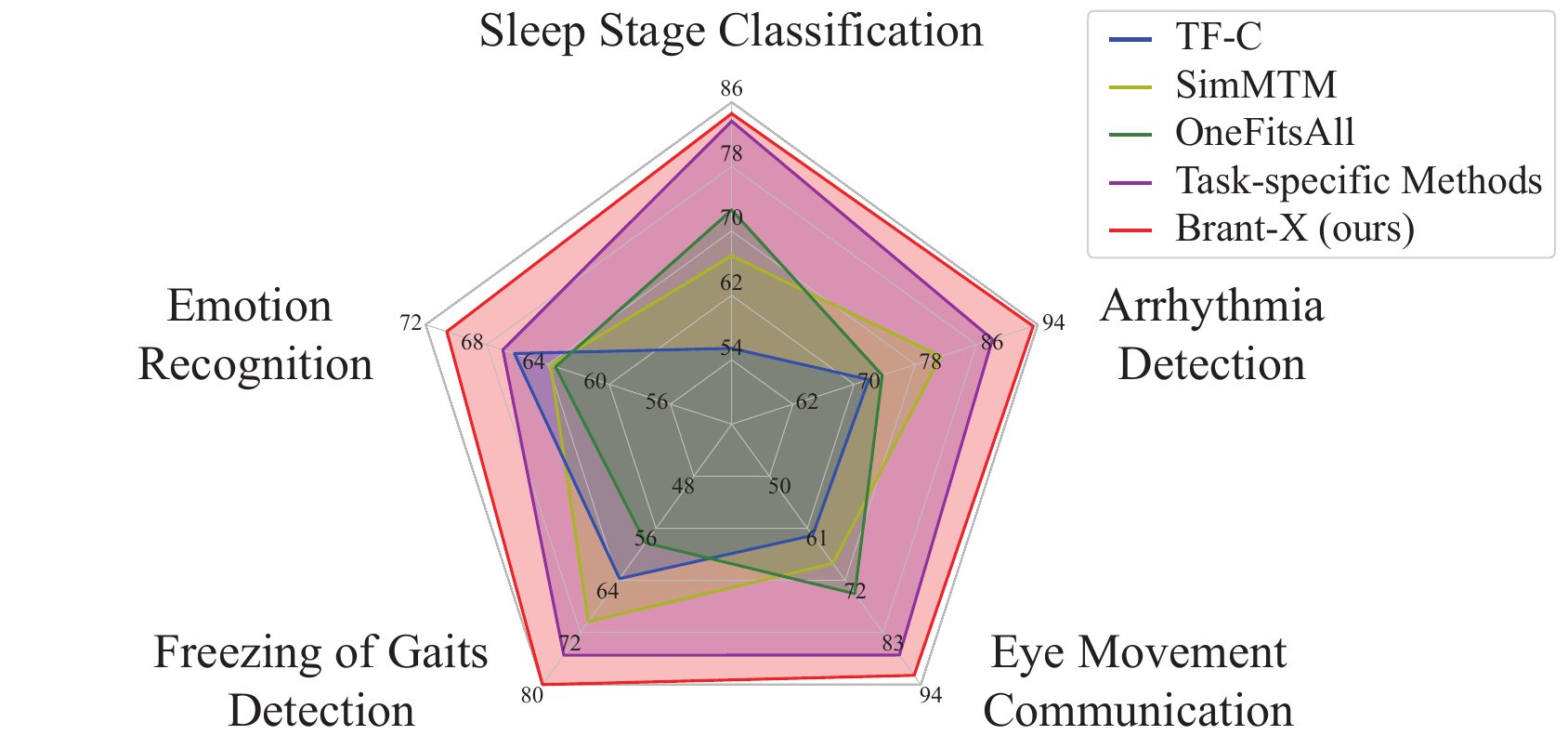}
  \caption{Overall performance comparison on various tasks. 
  \small 
  }
  \label{fig:radar}
  \vspace{-3mm}
\end{figure}

\subsection{Experimental Results} 

Fig.~\ref{fig:radar} summarizes the overall accuracy of \model and other baselines on various downstream tasks (including the arrhythmia detection in Sec.~\ref{sec:ablation}). 
Since the task-specific methods vary across different tasks, we use ``Task-specific Methods'' to collectively represent their best results on each task.
As shown in Fig.~\ref{fig:radar}, compared with other baseline methods, \model achieves SOTA performance on all of the five tasks, illustrating the effectiveness of our framework in various scenarios.
Detailed comparisons on each task are discussed in following paragraphs,
where in all the tables we mark values ranking the first (\textbf{v}), second (\underline{v}) and third (v*) in each column.

\def\one #1 {\textbf{#1}}
\def\two #1 {\underline{#1}~}
\def\thr #1 {#1*}
\def \s   #1 {\footnotesize $\pm$#1}

\begin{table*}[ht]
    \caption{Average performance on the emotion recognition task. } 
    \vspace{-2mm}
    \renewcommand{\arraystretch}{0.95}  
    \centering
    \setlength\tabcolsep{3.8pt}
    \label{tab:exp_emo}

    \begin{tabular}{llllllllll}
    \toprule
    \multicolumn{1}{c}{\multirow{2}{*}{\diagbox{Methods}{Metrics}}}
    & \multicolumn{3}{c}{Valence} & \multicolumn{3}{c}{Arousal}  & \multicolumn{3}{c}{Dominance} \\
    \cmidrule(lr){2-4} \cmidrule(lr){5-7} \cmidrule(lr){8-10} 
                             & \multicolumn{1}{c}{Acc.} & \multicolumn{1}{c}{F1} & \multicolumn{1}{c}{AUC} & \multicolumn{1}{c}{Acc.} & \multicolumn{1}{c}{F1} & \multicolumn{1}{c}{AUC} & \multicolumn{1}{c}{Acc.} & \multicolumn{1}{c}{F1} & \multicolumn{1}{c}{AUC}         \\
    \midrule 
    TF-C~\citep{zhang2022self}              & 66.20 \s3.76 & 78.09 \s5.02 & 69.71 \s7.34  & \thr76.45 \s11.36 & 85.86 \s8.23 & \two80.40 \s10.27 & 78.17 \s9.64  & 87.01 \s6.57  & 85.20 \s4.81  \\
    SimMTM~\citep{dong2023simmtm}           & 63.84 \s5.93 & 75.52 \s5.90 & 69.73 \s4.02  & 76.16 \s7.97  & \thr86.21 \s5.23 & 76.42 \s12.39 & 78.54 \s3.94  & 87.81 \s2.54  & 82.84 \s7.99  \\
    OneFitsAll~\citep{zhou2023onefitsall}   & 63.51 \s6.66 & 76.93 \s5.03 & 64.71 \s11.22 & 73.88 \s7.28  & 83.84 \s6.14 & 76.75 \s7.33  & 77.41 \s4.94  & 86.92 \s3.31  & \thr85.59 \s5.60  \\
    Time-LLM~\citep{jin2023time} & \two68.03 \s5.82          & 72.22 \s5.27          & \textbf{80.83 \s6.04} & 76.39 \s7.45          & 85.63 \s6.18          & 80.28 \s7.73          & 80.10 \s4.81          & 88.92 \s3.45          & 79.68 \s5.07          \\
    MiniRocket~\citep{dempster2021minirocket} & 60.54 \s7.09 & 65.68 \s6.80 & 64.36 \s8.05 & 75.73 \s8.89 & 85.75 \s7.64 & 77.90 \s10.46 & 75.11 \s5.91 & 85.28 \s5.14 & \two86.69 \s7.16 \\

    \midrule
    MLF-CapsNet~\citep{liu2020multi}        & 65.67 \s2.87 & 77.06 \s3.87 & 71.05 \s4.66  & 74.56 \s7.49  & 84.98 \s5.32 & 79.80 \s10.92 & 77.13 \s2.36  & 86.94 \s1.35  & 82.61 \s8.21  \\
    EEG Conformer~\citep{song2022eeg}       & 59.82 \s7.05 & 69.53 \s6.93 & \thr71.94 \s11.91 & 73.07 \s9.67  & 83.21 \s7.41 & 75.11 \s7.65  & \thr81.82 \s6.05  & \thr89.50 \s4.13  & 83.19 \s9.87  \\
    \citet{lin2023eeg}                      & 66.47 \s6.85 & \thr79.50 \s5.04 & 67.10 \s8.94  & 75.54 \s7.81  & 85.87 \s5.14 & 79.06 \s6.65  & 78.46 \s5.04  & 87.83 \s3.12  & 79.40 \s6.50  \\
    \citet{wang2023novel}                   & \thr66.95 \s8.30 & \two79.84 \s6.11 & 66.20 \s10.73  & \two76.47 \s9.29  & \two86.44 \s6.14 & \thr80.29 \s7.58  & \two81.87 \s5.31  & \two89.96 \s3.22  & 83.97 \s5.81  \\
    \midrule
    \model                  & \one70.61 \s4.01 & \one80.51 \s3.81 & \two72.48 \s4.10  & \one78.64 \s8.56  & \one87.59 \s5.71 & \one82.14 \s7.98  & \one83.54 \s5.27  & \one90.97 \s3.16  & \one90.19 \s4.94  \\
    \bottomrule
    \end{tabular}
    
    \vspace{-2mm}
\end{table*}

\def\one #1 {\textbf{#1}}
\def\two #1 {\underline{#1}~}
\def\thr #1 {#1*}
\def \s   #1 {\footnotesize $\pm$#1}

\begin{table}[ht]
  \caption{Average performance on the FoG detection task.}
  \vspace{-2mm}
  \renewcommand{\arraystretch}{0.95}  
  \label{tab:exp_fog}
   \setlength\tabcolsep{3.2pt}

    \begin{tabular}{lllll}
    \toprule
    Methods & \multicolumn{1}{c}{Acc.} & \multicolumn{1}{c}{Prec.} & \multicolumn{1}{c}{Rec.} & \multicolumn{1}{c}{F1}         \\
    \midrule
    TF-C~\citep{zhang2022self}              & 63.72 \s1.83 & 61.28 \s2.91 & \two76.14 \s6.02  & 67.77 \s2.52 \\
    SimMTM~\citep{dong2023simmtm}           & 70.32 \s4.22 & 69.05 \s8.65 & \thr74.79 \s7.98 & 71.88 \s0.20  \\
    OneFitsAll~\citep{zhou2023onefitsall}   & 58.22 \s3.31 & 56.62 \s4.13 & 71.80 \s16.41 & 62.98 \s4.19 \\
    Time-LLM~\citep{jin2023time} & 72.73 \s2.98 & \thr74.23 \s4.75 & 69.14 \s5.54 & 71.62 \s4.66    \\    
    MiniRocket~\citep{dempster2021minirocket} & 73.42 \s2.02 & 72.73 \s2.07 & 72.11 \s0.69 & 72.18 \s1.38 \\
    
    \midrule
    Aly et al.~\citep{aly2023bio}               & 72.12 \s3.31 & 71.09 \s3.41 & 73.54 \s5.10  & 72.24 \s3.60 \\
    Batool et al.~\citep{batool2022movement}    & \two75.49 \s2.35 & \two75.50 \s2.56 & 74.62 \s3.22  & \two75.04 \s2.59 \\
    \citet{goel2023ensemble}                    & \thr74.18 \s2.10 & 73.49 \s3.59 & 74.58 \s2.64  & \thr73.96 \s1.98 \\
    \midrule
    \model                    & \one80.14 \s1.33 & \one81.97 \s1.56 & \one76.73 \s3.80  & \one79.21 \s1.86 \\ 
    \bottomrule
    \end{tabular}

    \vspace{-3mm}
\end{table}

\def\one #1 {\textbf{#1}}
\def\two #1 {\underline{#1}~}
\def\thr #1 {#1*}
\def \s   #1 {\footnotesize $\pm$#1}

\begin{table}[ht]
  \caption{Average performance on the eye movement communication task.}
  \vspace{-2mm}
  \renewcommand{\arraystretch}{0.95}  
  \label{tab:exp_eye}
   \setlength\tabcolsep{1.7pt}

    \begin{tabular}{lllll}
    \toprule
    Methods      & \multicolumn{1}{c}{Acc.} & \multicolumn{1}{c}{Prec.} & \multicolumn{1}{c}{Rec.} & \multicolumn{1}{c}{F1}        \\
    \midrule
    TF-C~\citep{zhang2022self}              & 62.50 \s4.52 & 61.90 \s4.18 & 65.00 \s9.61  & 63.41 \s3.94 \\
    SimMTM~\citep{dong2023simmtm}           & 68.42 \s5.42 & 74.95 \s4.16  & 56.06 \s10.62  & 63.83 \s8.54 \\
    OneFitsAll~\citep{zhou2023onefitsall}   & 74.81 \s3.88 & 75.90 \s3.72 & 72.69 \s8.64 & 74.26 \s3.61 \\
    Time-LLM~\citep{jin2023time} & 81.78 \s4.37 & 84.71 \s4.65 & 77.54 \s7.36 & 80.96 \s6.14    \\
    MiniRocket~\citep{dempster2021minirocket} & 71.43 \s5.37 & 63.64 \s6.40 & \two93.31 \s8.63 & 75.68 \s6.07 \\

    \midrule
    eyeSay~\citep{zou2021eyesay}            & 80.24 \s6.61  & 84.43 \s4.66  & 74.75 \s9.81  & 79.18 \s7.35 \\
    Adama et al.~\citep{adama2021yes}       & \two87.75 \s8.41  & \two87.86 \s7.30  & \thr87.41 \s10.71 & \two87.56 \s8.70 \\
    \citet{hossieny2022developing}          & \thr83.34 \s5.19  & \thr86.33 \s4.74  & 79.02 \s7.04  & \thr82.47 \s5.77 \\
    \midrule
    \model      & \one92.04 \s3.13  & \one90.99 \s3.96  & \one93.42 \s2.87  & \one92.17 \s3.06 \\
    \bottomrule
    \end{tabular}

    \vspace{-3mm}
\end{table}
 
The performance comparison on sleep stage classification task is given in Tab.~\ref{tab:exp_sleep}. It shows that \model achieves top rankings
in almost all performance metrics, demonstrating that \model can effectively transfer the knowledge from EEG to EOG signals, combining the information of both EEG and EOG signals to learn the high-level semantic information therein. 
The baselines on general time series did not yield good results, mainly because these models struggle 
to overcome the huge gap in inherent features between various physiological signals, and do not model correlations from different semantic scales like \model does.

As shown in Tab.~\ref{tab:exp_emo}, on emotion recognition task, \model achieves SOTA performance compared to all the baselines. 
Compared to the baselines designed solely for EEG, \citet{wang2023novel} claims the second spot, because it adopt the same strategy as \model for combining the information of both EEG and ECG, thereby demonstrating stronger learning capabilities. 
However, \model still surpasses \citet{wang2023novel} on all metrics, benefiting from alignment training based on contrastive learning. 

\begin{figure*}
  \centering
  \includegraphics[width=\linewidth]{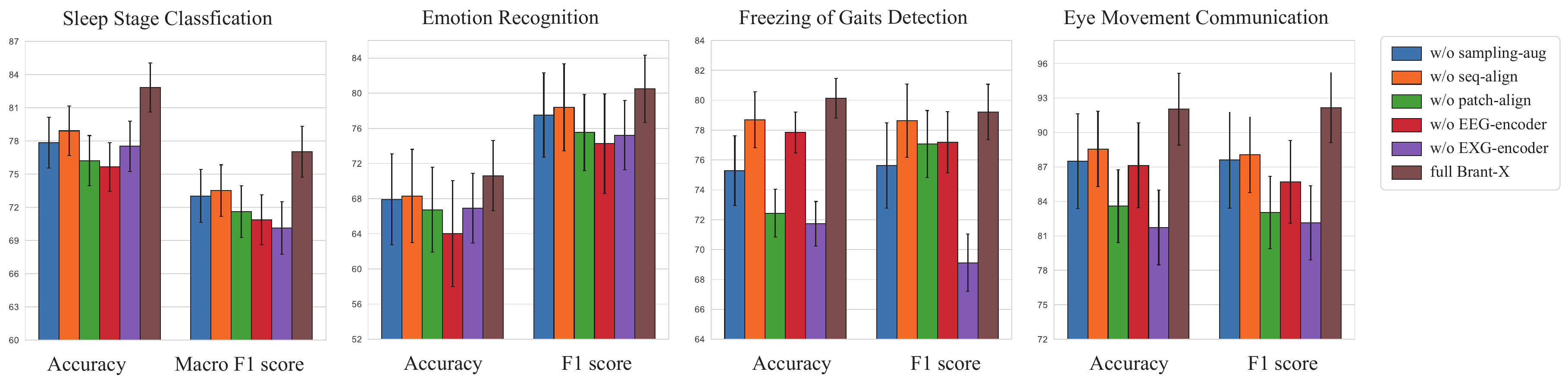}
  \vspace{-5mm}
  \caption{Results of the ablation study on all the downstream tasks.
  }
  \label{fig:ablation}
  \vspace{-1mm}
\end{figure*}

The overall results on the freezing of gaits detection and eye movement communication tasks are given in Tab.~\ref{tab:exp_fog} and Tab.~\ref{tab:exp_eye}, respectively. 
\model defeats all the baseline methods on these two tasks, showing its ability to learn representations from simultaneously collected EEG, EMG, and EOG data. 
Batool et al.~\citep{batool2022movement} and Adama et al.~\citep{adama2021yes} achieve the second-best performance on these two tasks, respectively, mainly because they explicitly extract the frequency domain information of physiological signals as inherent features for physiological data modeling.

\vspace{-2mm}

\subsection{Ablation Study} \label{sec:ablation}

To evaluate the effectiveness of each component in \model, we conduct ablation experiments on four model variants, including: 
(1) \textit{Brant-X w/o sampling-aug}: \model without the sampling augmentation during alignment; 
(2) \textit{Brant-X w/o patch-align}: \model without the patch-level alignment; 
(3) \textit{Brant-X w/o seq-align}: \model without the sequence-level alignment; 
(4) \textit{Brant-X w/o EEG-encoder}: \model without the EEG encoder during downstream evaluation after alignment;
(5) \textit{Brant-X w/o EXG-encoder}: \model without the EXG encoder during downstream evaluation after alignment. 

The comparison results of the ablation experiments on the four downstream tasks are presented in Fig.~\ref{fig:ablation}. 
It demonstrates that \model outperforms other variants on all metrics of all the tasks, evidencing the contribution of each component in our framework. 
Compared to the full \model, the performance of \textit{Brant-X w/o sampling-aug} decreases, showing the boost of model robustness against variable sampling rates provided by the sampling augmentation. 
Also, \textit{Brant-X w/o patch-align} and \textit{Brant-X w/o seq-align} show a decrease in performance, suggesting that the two-level alignment can align EEG and EXG signals from different semantic scales to learn informative representations from physiological data. 
For sleep staging and emotion recognition, \textit{Brant-X w/o EEG-encoder} drops greatly in performance, as EEG signals play an important role in these scenarios. This corroborates the significance of the brain as a central control in vital activities, as we emphasized in Sec.~\ref{sec:intro}. 

\vpara{EXG Encoder Analysis. } 
As a supplement to the \textit{Brant-X w/o EEG-encoder} in the ablation experiments, we extend our assessment to more tasks using the standalone EXG encoder, to validate whether the EXG encoder can learn useful representations from EXG data during the alignment training. 
Specifically, we conduct experiments with the aligned EXG encoder on ECG data (without incorporating the EEG encoder on EEG data) on the arrhythmia detection task. 
More details about this task and the results are given in  App.~\ref{app:arrhythmia}. 
As shown in Tab.~\ref{tab:exp_arrhythmia}, \model achieves SOTA performance on the arrhythmia detection task, showing that the alignment training indeed enables the EXG encoder to learn the representations from ECG signals, and then effectively classify cardiac rhythms.

\subsection{Case Study} 

\begin{figure}[h]
  \centering
  \includegraphics[width=\linewidth / 10 * 10]{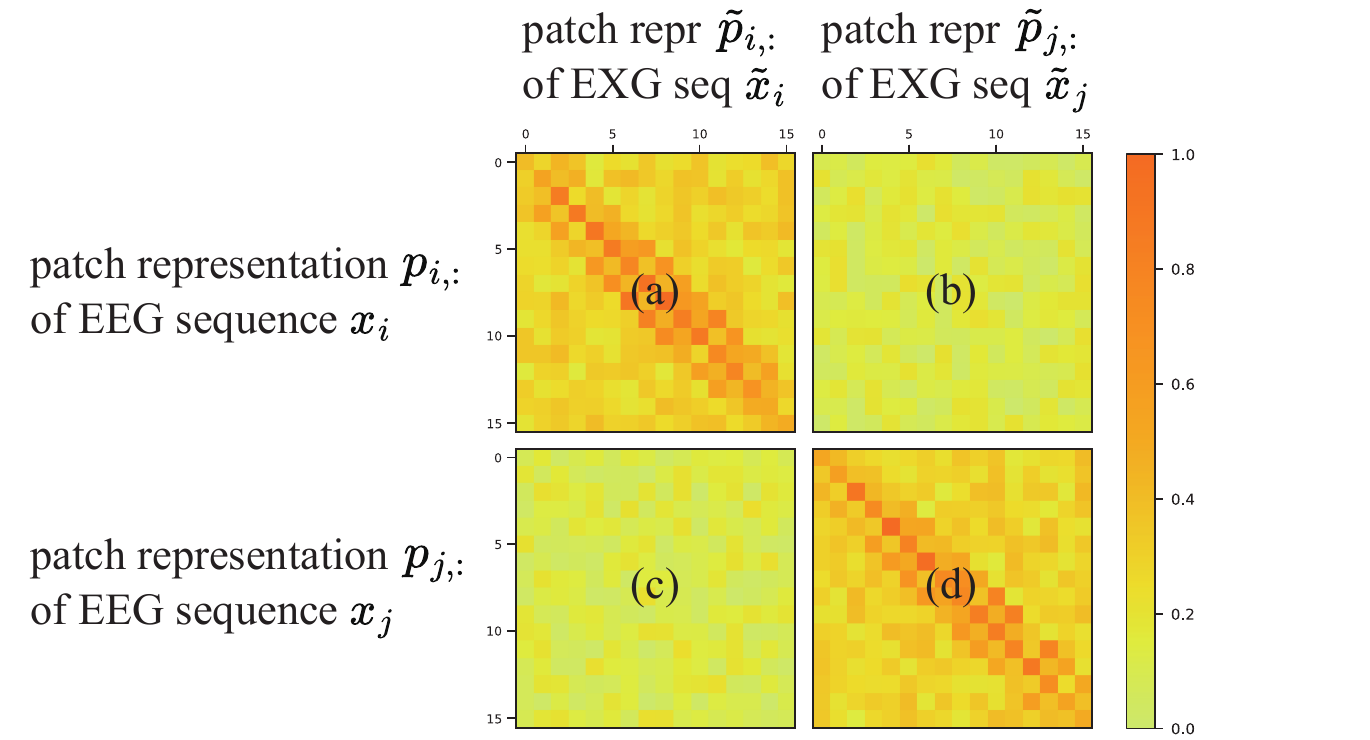}
  \vspace{-4mm} 
  \caption{Similarity matrices of patch representations. 
  \small The vertical axis represents the patch representations of two EEG sequences, $\boldsymbol{x}_i$ and $\boldsymbol{x}_j$, and the horizontal axis represents the patch representations of two EXG sequences, $\boldsymbol{\tilde{x}}_i$ and $\boldsymbol{\tilde{x}}_j$. 
  } 
  \label{fig:case}
  \vspace{-3mm}
\end{figure}

Fig.~\ref{fig:case} displays four similarity matrices between patch representations of two multi-type physiological data sequences, $\{ \boldsymbol{x}_i, \boldsymbol{\tilde{x}}_i \}$ and $\{ \boldsymbol{x}_j, \boldsymbol{\tilde{x}}_j \}$. 
The vertical axis represents the patch representations of two EEG sequences, $\boldsymbol{x}_i$ and $\boldsymbol{x}_j$, and the horizontal axis represents the patch representations of two EXG sequences, $\boldsymbol{\tilde{x}}_i$ and $\boldsymbol{\tilde{x}}_j$. Thus, four similarity matrices are given in Fig.~\ref{fig:case}. The darker the colour of each small square, the higher the normalised similarity between the representations of two corresponding patches.

Among these, matrix (a) (or (d)) indicates the similarity of patch representations of simultaneously collected EEG sequence $\boldsymbol{x}_i$ (or $\boldsymbol{x}_j$) and EXG sequence $\boldsymbol{\tilde{x}}_i$ (or $\boldsymbol{\tilde{x}}_j$). 
It presents an overall darker colour, demonstrating the correlations between patches from the simultaneously collected EEG and EXG sequence. 
Moreover, the diagonal of matrix (a) (or (d)) is particularly dark, indicating that the simultaneous EEG and EXG patches are well-aligned.
However, as for matrices (b) and (c), they have an overall lighter colour, suggesting little to no correlation between patch representations of non-simultaneously collected EEG and EXG data.
These four similarity matrices in this case illustrate well that the two-level alignment can bring the representations of simultaneous EEG and EXG data closer, while distancing irrelevant sequences, such that \model can perform knowledge transfer from EEG to EXG.

\section{Related Work}

\vpara{Physiological Signal Modeling. }
With the maturation of physiological recording technology and the advancement of machine learning methods, physiological signal modeling has captivated many researchers. 
Initially, researchers mainly focus on model learning on a single type of signal. 
A large body of works propose to use time series~\citep{supratak2020tinysleepnet,phan2021xsleepnet,liu2020multi,song2022eeg,yuan2023ppi,gong2023astdf,liu2022bstt} or graph~\citep{chen2022brainnet,cai2023mbrain,lin2023eeg} data structures with supervised~\citep{supratak2020tinysleepnet,phan2021xsleepnet,liu2020multi,song2022eeg,lin2023eeg,gong2023astdf,liu2022bstt} or self-supervised~\citep{chen2022brainnet,cai2023mbrain,yuan2023ppi} learning paradigms for various tasks on EEG signals. 
Recently, some large EEG models~\citep{jiang2024large,zhang2023brant,yuan2024brant} also emerged, which break through the limitations of different tasks on EEG.
Also, methods based on feature engineering or supervised learning are proposed to learn representations from EOG~\citep{zou2021eyesay,adama2021yes,hossieny2022developing}, ECG~\citep{midani2023deeparr,alamatsaz2024lightweight}, and EMG~\citep{montazerin2022vit,zabihi2023trahgr} signals. 
Additionally, to fully mine the potential semantics of physiological data, research attention has been drawn to the modeling of multi-type signals. 
\citet{jia2023exploiting} consider the interactivity of EEG, EOG and EMG signals, imporving the SOTA performance on sleep staging task. 
\citet{wang2023novel} fuse the features from single-lead EEG and ECG data for emotion recognition. 
\citet{aly2023bio} propose an approach that integrates EEG with EMG signals, boosting the performance of hand and wrist motion control. 
However, most works on physiological signals are task-specific or signal-specific. 
They neither leverage foundation models (thus are hindered by the data scarcity), nor possess a unified framework for various physiological signals across a range of tasks.

\vpara{Multimodal Alignment. }
To capitalize on the information consistency in multimodal data, contrastive-learning-based alignment strategy has achieved impressive results in many fields like image-text~\citep{radford2021learning,yang2022unified,yu2022coca}.  
For physiological research, 
methods~\citep{chen2020unsupervised,chaitanya2020contrastive} conduct alignment to the features of images for medical images segmentation. 
\citet{wang2022multi} aligns the paired medical image and radiology reports (text) for image classification and object detection, etc.
\citet{fan2022unsupervised} propose a domain adaptation approach to bridge the gap between the EEG data distribution of source and target domains for sleep staging. 
\citet{lv2021progressive} reinforce features by aligning the visual and acoustic modality within video clips for emotion recognition. 
However, these studies primarily explore the consistency among text, image, or audio modalities,  none of which explicitly align the simultaneously collected physiological signals. 

\vpara{Time seires Modeling. }
Time series (TS) analysis has been utilized in many real-world applications, including finance, meteorology, healthcare, and so on, attracting more and more researchers.
\citet{wu2022timesnet} propose TimesNet as a task-general backbone to discover the multi-periodicity adaptively for TS analysis.
\citet{dong2023simmtm} propose to recover masked time points by the weighted aggregation of multiple neighbors outside the manifold for TS modeling. 
\citet{zhou2023onefitsall} propose a unified model that leverages language or vision models for TS analysis. 
\citet{jin2023time} present a reprogramming framework named Time-LLM to repurpose large language models for general TS forecasting. 
However, most TS works cannot adapt well to high-frequency physiological signals and ignore the correlation between physiological signals.
\section{Conclusion}

In this work, we are the first to propose a unified physiological signal alignment framework, \model.
Based on the EEG foundation model, we summarize available multi-type physiological datasets, to transfer the 
rich knowledge from the EEG foundation model to EXG signals.
We adopt the two-level alignment that aligns the semantics of EEG and EXG data at both patch- and sequence-level, to adapt to various downstream scenarios.  
In this way, EEG is viewed as a bridge between the EEG foundation model and EXG data, allowing the data and model resources in the EEG field to empower the research on other physiological signals, paving a new avenue to model the correlations between various physiological signals. 
In the future, motivated by the positive results of \model, it would be intriguing to explore further studies along this research line on more physiological signals.  

\vpara{Acknowledgment.} This work is supported by National Natural Science Foundation of China (No. 62322606, No. 62441605) and SMP-IDATA Open Youth Fund.

\bibliographystyle{ACM-Reference-Format}
\bibliography{reference}

\appendix

\section{Details of the EXG Encoder}\label{app:exg_encoder}

Because the focus of this paper is to introduce our proposed alignment framework, the specific encoder architecture can be flexible. 
The model architecture of the EXG encoder used in this paper are introduced here. 

Since physiological signals are bioelectric signals, the time domain provides information about the amplitude and duration, while the frequency domain can reveal the oscillation patterns and underlying biological rhythms~\citep{kirby2023time}. Therefore, to combine the information from both time and frequency domains, in our EXG encoder, we first calculate the power spectral density (PSD)~\citep{stoica2005spectral}, which describes the distribution of the signal’s total average power over frequency, as the information in frequency domain. 
Then, a convolutional neural network (CNN) performs on the PSD to extract features in the frequency domain of the EXG signal. 
The extracted features in the frequency domain will be concatenated with the convolution-derived features in the time domain, serving as the features within a single patch. 
Due to the fact that physiological signals are time series, each patch has a temporal dependency with its contextual patches from the same sequence. 
With this in mind, the features of consecutive patches from a sequence will be fed into the Transformer~\citep{vaswani2017attention} to obtain a more comprehensive representation that considers temporal dependencies. 

Formally, all the patches $\{ \boldsymbol{\tilde{x}}_{i,j} \}_{j=0}^{P-1}$ from the $i$-th EXG data sequence $\boldsymbol{\tilde{x}}_i$ are input into the EXG encoder, generating their representations $\{ \boldsymbol{\tilde{p}}_{i,j} \}_{j=0}^{P-1}$: 
\begin{multline}
    \boldsymbol{\tilde{p}}_{i,j} = 
    \text{Transformer} \Big(
        \text{CNN}_T \left(\boldsymbol{\tilde{x}}_{i,j} \right)
        \mathbin\Vert
        \text{CNN}_F \left(\text{PSD} (\boldsymbol{\tilde{x}}_{i,j}) \right)
    \Big), \\ 
    j=0,1,2,...,P-1 ,
\end{multline}
where $\boldsymbol{\tilde{p}}_{i,j} \in \mathbb{R}^{D_p}$ denotes the representation of the $j$-th patch from the $i$-th EXG data sequence $\boldsymbol{\tilde{x}_i}$, and $D_p$ denotes the dimension of patch representations.

\section{Details of Baselines}\label{app:baseline}

Firstly, we compare \model with the existing self-supervised or unsupervised works on genral time series. The Details of these baseline models are given here: 
\begin{itemize}[leftmargin=*]
    \item TF-C~\citep{zhang2022self}: 
    A decomposable pre-training model for general time series modeling, where the self-supervised signal is provided by the distance between time and frequency components.

    \item SimMTM~\citep{dong2023simmtm}:
    A pre-training framework on time series to recover masked time points by the weighted aggregation of multiple neighbors outside the manifold.

\end{itemize}
Also, we compare \model with the methods that performs time series classification based on pre-trained language models. Hence, we set OneFitsAll~\citep{zhou2023onefitsall} as a baseline: 
\begin{itemize}[leftmargin=*]
    \item OneFitsAll~\citep{zhou2023onefitsall}: 
    A unified model that leverages language or vision models for time series analysis, leading to a comparable or SOTA performance in all main time series analysis tasks. 

\end{itemize}
Furthermore, to illustrate the effectiveness of \model in various application scenarios, we compare our framework with the SOTA methods those are specially designed for each of the four downstream tasks. These supervised methods includes: 

\def\one #1 {\textbf{#1}}
\def\two #1 {\underline{#1}~}
\def\thr #1 {#1*}
\def \s   #1 {\footnotesize $\pm$#1}
\begin{table*}[ht]
    \caption{Average performance on the arrhythmia detection task. }
    
    \setlength\tabcolsep{1.8pt}
    \centering
    \renewcommand{\arraystretch}{0.95}  
    \label{tab:exp_arrhythmia}

    \begin{tabular}{lllllllllll}
    \toprule
    \multicolumn{1}{c}{\multirow{2}{*}{\diagbox{Methods}{Metrics}}}
    & \multicolumn{1}{c}{Overall}    & \multicolumn{3}{c}{N rhythm}           & \multicolumn{3}{c}{A rhythm}           & \multicolumn{3}{c}{O rhythm}            \\
    \cmidrule(lr){2-2} \cmidrule(lr){3-5} \cmidrule(lr){6-8} \cmidrule(lr){9-11}    
                      & \multicolumn{1}{c}{Acc.} & \multicolumn{1}{c}{Sens.} & \multicolumn{1}{c}{Spec.} & \multicolumn{1}{c}{Prec.} & \multicolumn{1}{c}{Sens.} & \multicolumn{1}{c}{Spec.} & \multicolumn{1}{c}{Prec.} & \multicolumn{1}{c}{Sens.} & \multicolumn{1}{c}{Spec.} & \multicolumn{1}{c}{Prec.}        \\
    \midrule
    TF-C~\citep{zhang2022self}              & 71.91 \s2.25 & 81.44 \s10.07 & 25.37 \s10.91 & 64.61 \s0.67 & 3.51 \s2.15   & 96.88 \s0.99 & 8.89 \s2.92   & 22.95 \s9.85  & 84.08 \s9.42  & 39.33 \s6.10  \\
    SimMTM~\citep{dong2023simmtm}           & 81.30 \s2.57 & 83.60 \s10.91 & 65.43 \s15.21 & 81.27 \s4.97 & \thr59.13 \s10.14 & 95.01 \s3.28 & 56.27 \s12.18 & 49.58 \s14.21 & 85.00 \s8.18 & 58.13 \s8.02    \\
    OneFitsAll~\citep{zhou2023onefitsall}   & 73.67 \s1.92 & 83.88 \s11.38 & 25.40 \s15.49 & 66.49 \s1.57 & 9.90 \s6.58   & 97.51 \s2.28 & 34.41 \s12.32 & 22.53 \s14.74 & 86.00 \s10.62 & 43.35 \s10.46 \\
    \midrule
    DeepArr~\citep{midani2023deeparr}       & \thr86.94 \s1.67 & \thr94.03 \s4.51  & \thr66.29 \s15.48 & \two83.89 \s5.55 & 52.09 \s19.37 & \thr98.20 \s1.72 & \thr77.34 \s11.17 & \thr57.08 \s19.78 & \thr91.89 \s6.49  & \thr76.81 \s12.13 \\
    \citet{alamatsaz2024lightweight}        & \two88.08 \s1.45 & \two94.97 \s2.29  & \two67.66 \s4.92  & \thr83.59 \s1.92 & \two60.28 \s11.09 & \two98.66 \s0.76 & \two81.59 \s8.83  & \two59.81 \s4.26  & \two93.33 \s1.93  & \two77.94 \s5.18  \\
    \midrule
    \model                                  & \one93.40 \s1.63 & \one96.46 \s3.14  & \one83.28 \s3.59  & \one90.96 \s1.73 & \one79.83 \s7.29  & \one99.62 \s0.23 & \one95.19 \s2.55  & \one78.80 \s4.87  & \one95.21 \s3.44  & \one87.14 \s7.42  \\ 
    \bottomrule
    \end{tabular}

\end{table*}

(1) For the sleep stage classification task:
\begin{itemize}[leftmargin=*]
    \item TinySleepNet~\citep{supratak2020tinysleepnet}:
    An end-to-end model based on CNN and LSTM for automatic sleep stage scoring on raw single-channel EEG with a less number of trainable parameters. 

    \item XSleepNet~\citep{phan2021xsleepnet}: 
    A sequence-to-sequence sleep staging model that is capable of learning a joint representation from both raw signals and time-frequency images. 

    \item L-SeqSleepNet~\citep{phan2023seqsleepnet}: 
    A method for efficient long sequence modelling that considers whole-cycle sleep information for sleep staging, showing robustness in alleviating classification errors.

    \item SleepHGNN~\citep{jia2023exploiting}: 
    A novel sleep heterogeneous graph neural network designed to capture interactivity and heterogeneity of physiological signals for accurate sleep stage classification. 
    
\end{itemize}

(2) For the emotion recognition task:
\begin{itemize}[leftmargin=*]
    \item MLF-CapsNet~\citep{liu2020multi}: 
    A multi-level features guided capsule network for multi-channel EEG-based emotion recognition, which can simultaneously extract features from the raw EEG signals and determine the emotional states. 

    \item EEG-Conformer~\citep{song2022eeg}: 
    A compact convolutional Transformer to encapsulate local and global features in a unified EEG classification framework for motor imagery and emotion recognition. 

    \item \citet{lin2023eeg}: 
    A graph convolution model with dynamic channel selection for emotion classification, which combines the advantages of 1D convolution and graph convolution to capture the intra- and inter-channel EEG features. 

    \item \citet{wang2023novel}: 
    An emotion recognition method based on the feature fusion of single-lead EEG and ECG signals, using various time-domain, frequency-domain, and nonlinear features. 
\end{itemize}

(3) For the freezing of gaits detection task:
\begin{itemize}[leftmargin=*]
    \item \citet{aly2023bio}: 
    A model based on CNN and LSTM that integrates EEG with EMG signals to investigate the efficiency of deep learning in hybrid systems with signal fusion for motion classification. 

    \item \citet{batool2022movement}:
    A feature engineering method that uses time-frequency feature extraction strategy and CNN-BiLSTM to detect walking disorder in Parkinson's disease patients. 

    \item \citet{goel2023ensemble}:
    An ensemble techniques that combines the prediction of multiple methods to improve the model performance for freezing of gaits detection on EEG signals
    
\end{itemize}

(4) For the eye movement communication task:
\begin{itemize}[leftmargin=*]
    \item eyeSay~\citep{zou2021eyesay}: 
    A multi-stage convolutional neural network to decode eye dynamics using electrooculography, towards voice-free communication for patients with amyotrophic lateral sclerosis. 

    \item \citet{adama2021yes}:
    A feature engineering method that employs features like relative power, spectral edge frequencies and symbolic mutual information for eye movement classification. 

    \item \citet{hossieny2022developing}:
    A model based on ResNet\citep{he2016deep} using horizontal and vertical EOG signals to determine six eye movement directions. 
\end{itemize}

For some baselines that are not open source, we re-implemented them for experiments. 
In order to make a fair comparison, for baselines designed for only one type of physiological signal (EEG or EXG), we take their best results on the following three settings as their final results: only on EEG, only on EXG, and aggregation the representation from EEG and EXG.

\section{Analysis on the Arrhythmia Detection Task} \label{app:arrhythmia} 

\textit{Atrial fibrillation} (AF) is the most common sustained cardiac arrhythmia, occurring in about 2\% of the general population and is associated with significant mortality and morbidity through association of risk of death, stroke, heart failure and coronary artery disease~\citep{clifford2017af}. Therefore, accurate rhythm classification and arrhythmia detection are vital to the prevention and treatment of heart disease. 
Depending on the different classifications of cardiac states, the task can be viewed as a multi-classification problem. 

The AFDB dataset~\citep{clifford2017af} comprises 12,186 single lead ECG recordings of 30 and 60sec long, gathered from subjects undergoing long-haul mobile ECG checking. Data are collected at 300Hz, and each sample may belong to one of four classes: (1) normal sinus rhythm, (2) AF, (3) other rhythm, or (4) too noisy to classify. In our experiment, we remove the noisy samples so that this task is a three-class classification problem. 

As a supplement to \textit{Brant-X w/o EEG-encoder} in the ablation study, we conduct this experiment with the aligned EXG encoder on ECG data (without incorporating the EEG encoder on EEG data). 
The experiment is conducted on training, validation and test data in a 3:1:1 ratio and repeated to obtain the overall results. For each of the above three classes, we use sensitivity, specificity and precision as metrics to evaluate the performance of our aligned EXG encoder and other baselines. We also report the overall accuracy as an overall assessment of model performance. 
In line with the main experiments on the four main tasks, besides TF-C~\citep{zhang2022self}, SimMTM~\citep{dong2023simmtm} and OneFitsAll~\citep{zhou2023onefitsall}, we also compare our EXG encoder with the SOTA methods in the field of arrhythmia detection to demonstrate the effectiveness of our model. These methods includes DeepArr~\citep{midani2023deeparr} and ~\citet{alamatsaz2024lightweight}. 

As shown in Tab.~\ref{tab:exp_arrhythmia}, our \model beats all of the baselines on the arrhythmia detection task. This demonstrates that the phase of alignment training empowers the EXG encoder to effectively learn semantic representations from ECG and classify cardiac rhythms.

\end{document}